\documentclass[final]{pro12}

\usepackage[ansinew]{inputenc}
\usepackage{amsmath}
\usepackage{graphicx} 	

\usepackage{lineno}

\head{C. Lonsdale et al.} {Solar imaging using low frequency arrays}	

\begin{document}

\title{SOLAR IMAGING USING LOW FREQUENCY ARRAYS}	
\author{C. Lonsdale\adress{\textsl MIT Haystack Observatory, Westford, MA, USA}$\,$, L. Benkevitch$^\ast$, I. Cairns\adress{\textsl University of Sydney, Sydney, Australia}$\,$, M. Crowley\adress{\textsl University of Massachusetts, Boston, USA}$\,$, P. Erickson$^\ast$,\\ M. Knapp\adress{\textsl MIT, Cambridge MA, USA}$\,$, K. Kozarev\adress{\textsl Institute of Astronomy, Sofia, Bulgaria}$\,$, F. Lind$^\ast$, P. McCauley$^\dag$, J. Morgan\adress{\textsl Curtin University, Perth, Australia}$\,$,\\ and D. Oberoi\adress{\textsl NCRA-TIFR, Pune, India}}


\maketitle

\begin{abstract}
Low frequency imaging radio arrays such as MWA, LWA and LOFAR have been recently commissioned, and significantly more advanced and flexible arrays are planned for the near term. These powerful instruments offer new opportunities for direct solar imaging at high time and frequency resolution. They can also probe large volumes of the heliosphere simultaneously, by virtue of very large fields of view. They allow highly detailed, spatially resolved study of solar and heliospheric radio bursts, which are complemented by heliospheric propagation studies using both background astronomical radio emissions as well as the bursts themselves. In this paper, the state of the art in such wide field solar and heliospheric radio studies is summarized, including recent results from the Murchison Widefield Array (MWA). The prospects for major advances in observational capabilities in the near future are reviewed, with particular emphasis on the RAPID system developed at Haystack Observatory.
\end{abstract}

\section{Introduction}


The Sun is an inherently challenging object to image at low radio frequencies, exhibiting extreme variability in both time and frequency. Energetic particles and strong magnetic fields in the corona give rise to a wide variety of intense non-thermal radio emissions with a broad range of spectral properties, and timescales ranging from milliseconds to days. This creates difficulties for traditional aperture synthesis interferometric radio imaging, because one can use neither Earth rotation nor multifrequency synthesis to improve the sampling of the Fourier plan (also known as the $uv$ plane) and facilitate high  fidelity imaging.  Most existing instruments of this type rely heavily on such techniques, and are therefore not particularly well suited to solar imaging applications.

In this paper, we discuss a new generation of low frequency radio imaging arrays. These instruments exploit rapidly expanding data transport and signal processing capacities of modern devices to implement array layouts and correlation architectures that allow high fidelity imaging, at arbitrarily high spectral and temporal resolution.  They offer unprecedented instantaneous monochromatic $uv$ coverage, well suited to detailed study of targets like the Sun, which due to the signal strength, lead to high signal to noise ratios even within small temporal and spectral integration windows.  We focus on two such arrays; the MWA which is fully operational and which represents the current state of the art in low frequency solar imaging; and the RAPID array now in the final stages of development which promises to  significantly advance that state of the art for a broad range of strong signal scientific investigations.

\section{The Murchison Widefield Array (MWA)}

\subsection{The instrument}

The MWA was conceived in 2004 and developed over several years, culminating in formal commissioning and the start of operations in November 2012 [Lonsdale et al., 2009; Tingay et al., 2013], supporting a wide variety of science investigations [Bowman et al., 2013]. It comprises 128 antenna tiles, each containing 16 dual-polarization dipole antennas laid out in a 4x4 grid. Each tile includes an analog time-delay beamformer allowing the beam to be electronically steered in any direction, down to about 30 degrees elevation. The analog signal path covers the 80--300~MHz frequency range, and the field of view of the tiles is roughly $30 \times (100$~MHz$/\nu)$ degrees across. There are 16 digital receiver units, each serving 8 tiles. The receivers transmit a flexibly selected 30~MHz subset of the full 80--300~MHz band, in the form of digital voltage samples, to a central location for correlation and further processing.

The tiles are laid out over a region 3~km in diameter in outback Western Australia, with a strong central concentration. Each tile signal is correlated independently, yielding  8128 simultaneous baselines and very dense instantaneous $uv$ coverage that supports high precision snapshot imaging. The array is depicted in Figure~1.

\begin{figure}[ht]
\centering
\includegraphics[width=1.0\textwidth]{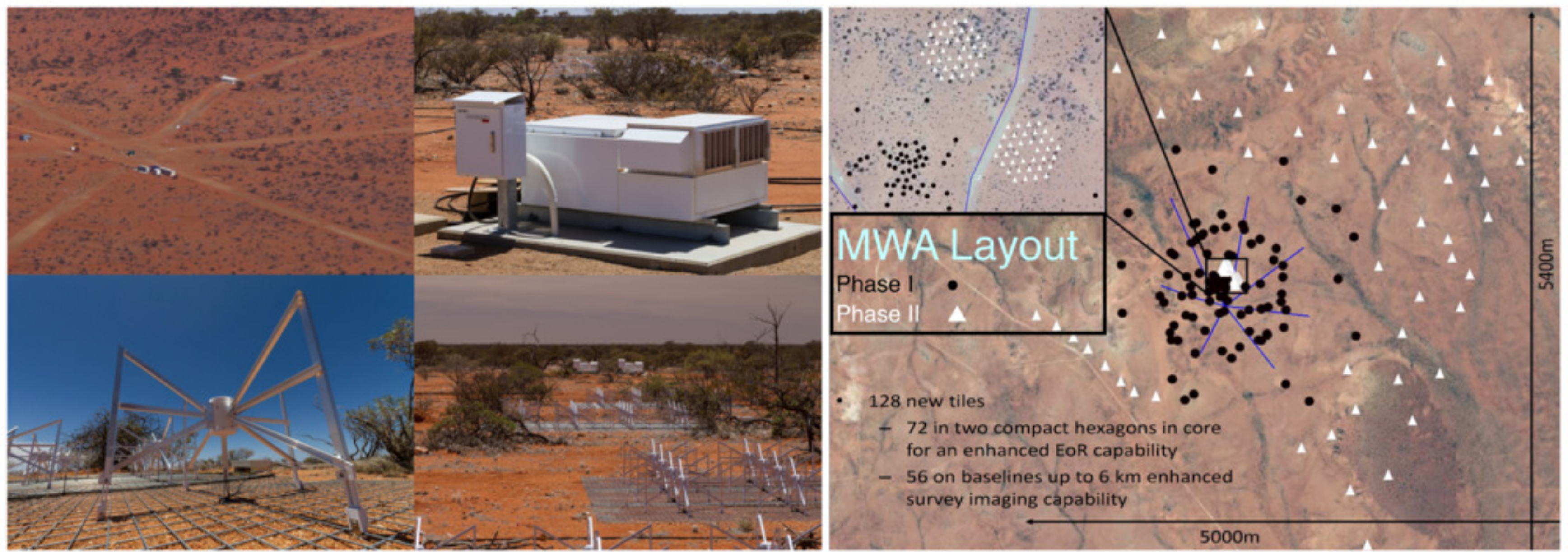}
\caption{The Murchison Widefield Array (MWA). The left panel shows (clockwise from top left) an aerial view of the center of the array, with trenching lines visible, a digital receiver unit servicing 8 tiles, a number of 4x4 tiles, and an individual dual-polarization dipole. The right panel illustrates the layout of tiles on the ground. Black dots are the 128-tiles of the phase 1 array, and the white triangles are the partially completed phase 2 tiles.}
\label{fig1}
\end{figure}

Two of the more challenging aspects of low frequency arrays are the corrupting influence of the Earth's ionosphere on interferometric phases, and the variation of those phase effects over the inherently wide fields of view -- of order 1000 square degrees in the case of the MWA. The process of solving simultaneously for the ionospheric phase screen, the brightness distribution on the sky and the instrumental properties (e.g. antenna gains and power patterns) is much more complex than is typically the case for higher-frequency, narrow-field arrays. The MWA addresses this problem through a combination of a very large number of independent measurements (by virtue of the 8128 baselines), and advanced algorithms to fully utilize those measurements. An excellent example of this can be found in the GLEAM survey [Hurley--Walker et al., 2017].  MWA wide field data reduction generally uses the Measurement Equation formalism [e.g. Smirnov 2011 and references therein].

\subsection{Solar imaging with the MWA}

For solar imaging, the above described problems are significantly reduced. Because the Sun is so bright relative to other astronomical sources at these frequencies, one can ignore the entire sky except for the roughly 1 square degree surrounding the Sun, effectively turning the MWA into a narrow field instrument for this application.  Ionospheric gradients across this narrow field can be neglected, and in principle the array can be accurately calibrated by solving for a single complex gain associated with each antenna tile, using conventional ``self-calibration" algorithms developed for higher frequency instruments. Each complex gain captures the combined effect of the ionosphere over the tile and the properties of the tile and signal path, in the direction of the Sun.

Of particular interest for solar and heliospheric research using the MWA is achieving high imaging dynamic range. This not only permits study of features such as gyrosynchrotron emission from recently erupted CME plasmas [e.g. Bastian, 2007], but also allows for high precision subtraction of the solar emission, to reveal emission from the rest of the field of view. This can enable, for instance, interplanetary scintillation studies or measurement of Faraday rotation in the inner heliosphere during the daytime, using appropriate wide field imaging techniques on the solar-subtracted dataset.  It may be noted that  scattering and refraction effects in the solar corona can blur and distort the apparent brightness distribution, but this does not significantly affect our ability to generate an accurate model of the  apparent structure, and to subtract that model from the data.

Extensive effort has therefore been expended to understand and mitigate a variety of effects limiting  solar imaging dynamic range using the MWA. Among these are somewhat obscure properties of low-bit-depth cross correlation at high correlated signal strengths, known as non-linear van Vleck corrections [Benkevitch et al., 2016], and bandwidth decorrelation effects unique to the MWA due to the architecture of the data transport and correlation systems. Work continues on these problems, and as of the current writing, an imaging dynamic range of 38,000:1 has been achieved in a single snapshot observation of 0.5 seconds duration and 40~kHz bandwidth.

\subsection{Solar burst spot mapping}

In September 2014 the MWA was observing the Sun during a bright type II burst event. Type II bursts are caused by shocks driven into the interplanetary medium by coronal mass ejections (CMEs), leading to strong, spectrally structured emission at the fundamental and second harmonic of the local plasma frequency.  The array was configured to  capture a contiguous 30~MHz band, and every 4 minutes to switch to a different 30~MHz frequency range. Fortuitously, significant parts of both the fundamental and harmonic burst emission were recorded, for 4 minutes each, the former in the 75-105 MHz range and the latter in the 105-135 MHz range. The unique snapshot imaging capabilities of the array allow independent high fidelity images to be made for each  pixel of a high resolution dynamic spectrum (i.e. for each time slice and for each frequency channel). Furthermore, the positional registration between the images is very precise due to the use of a common, stable instrumental calibration solution.

\begin{figure}[ht]
\centering
\includegraphics[width=1.0\textwidth]{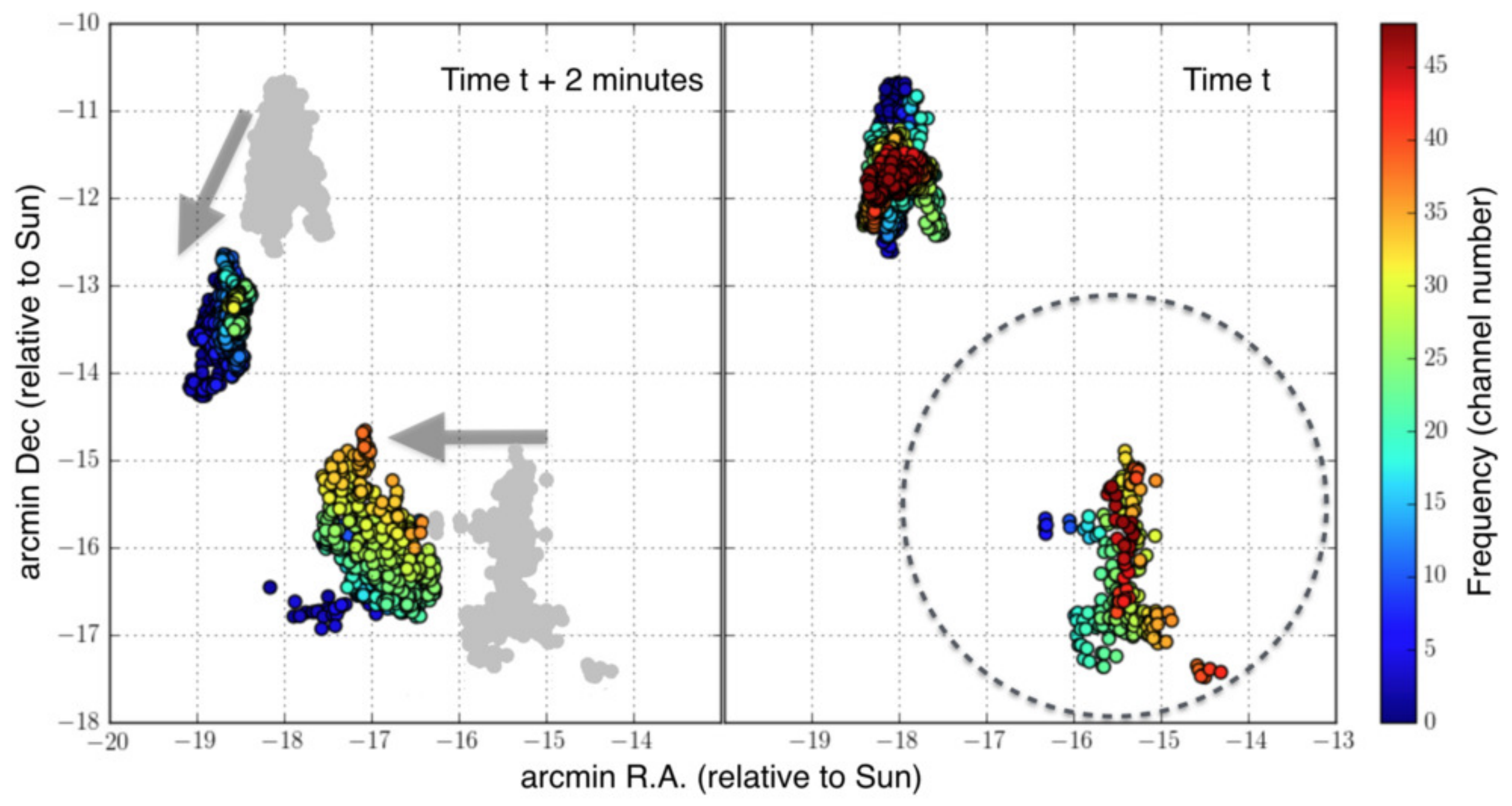}
\caption{Spot maps for the September 2014 type II burst event, observed with the MWA. In each panel, the lower right cluster of spots traces the fundamental emission in the 75-105 MHz frequency band, while the upper left cluster represents the harmonic emission 4 minutes later in the 105-135 MHz frequency band. Within each band, the color of the spots indicates the frequency of the emission split into 48 channels, and the left panel shows the change in position relative to the right panel over a period of 2 minutes, within each band. The dashed circle indicates the approximate angular resolution of the MWA at these frequencies, and the coherent fine-scale structures in the spot maps demonstrate the ability to extract spatial information on scale of a few percent of this resolution.}
\label{fig2}
\end{figure}

The type II emission in each image was barely resolved and was dominated by a single feature.  It is thus possible to  make accurate measurements of the emission centroid positions by gaussian peak fitting. This can be done on scales much smaller than the image resolution, since the nominal positional measurement error scales inversely with the signal to noise ratio, and the burst emission is very strong. By plotting these centroid positions for a range of times and frequencies, we can then construct ``spot maps" of the emission, and track positional dependencies with time and frequency with high precision.

A representation of these spot maps is shown in Figure~2, revealing a number of interesting features. The fundamental and harmonic emissions appear to originate in different locations, and both types of emission display a systematic motion outward from the Sun. The color coding (indicating frequency) allows us to trace regions of different local plasma frequency and therefore different electron densities. Some of these density structures appear quite complex. We also note that the spot maps display fine-scale structures, in some cases only a few percent of the size of the array resolution (dashed circle). The most straightforward interpretation of this observation is that at any given location on the dynamic spectrum, given sufficient time and frequency resolution, the emission typically consists of only a single compact spot. These data will be presented in more detail in a forthcoming paper [Oberoi et al., in preparation].

\section{The RAPID array}

\subsection{The instrument}

The Radio Array of Portable Interferometric Detectors (RAPID) is a  project funded by the US National Science Foundation and led by MIT Haystack Observatory, in collaboration with Cambridge University in the UK, and JPL. The central concepts of the array are the complete elimination of cabling for power or data transport, and engineering the system for logistically efficient transportation, deployment and reconfiguration. Each element in the array constitutes a fully independent data gathering instrument. Power cables are eliminated by solar panels and battery storage, while data transport is accomplished using on-board solid state disks and physical post-experiment upload at a central location, commonly a computing cluster running a cloud operating system at the investigator's home institution. Each element is highly integrated with a base that serves as support and ground plane for the antenna, a flat surface for mounting the solar panels, a housing for the electronics and power systems, and a stackable shipping box. A target for assembly and deployment of a single RAPID unit to an operational state is 20 person-minutes.

\begin{figure}[ht]
\centering
\includegraphics[width=1.0\textwidth]{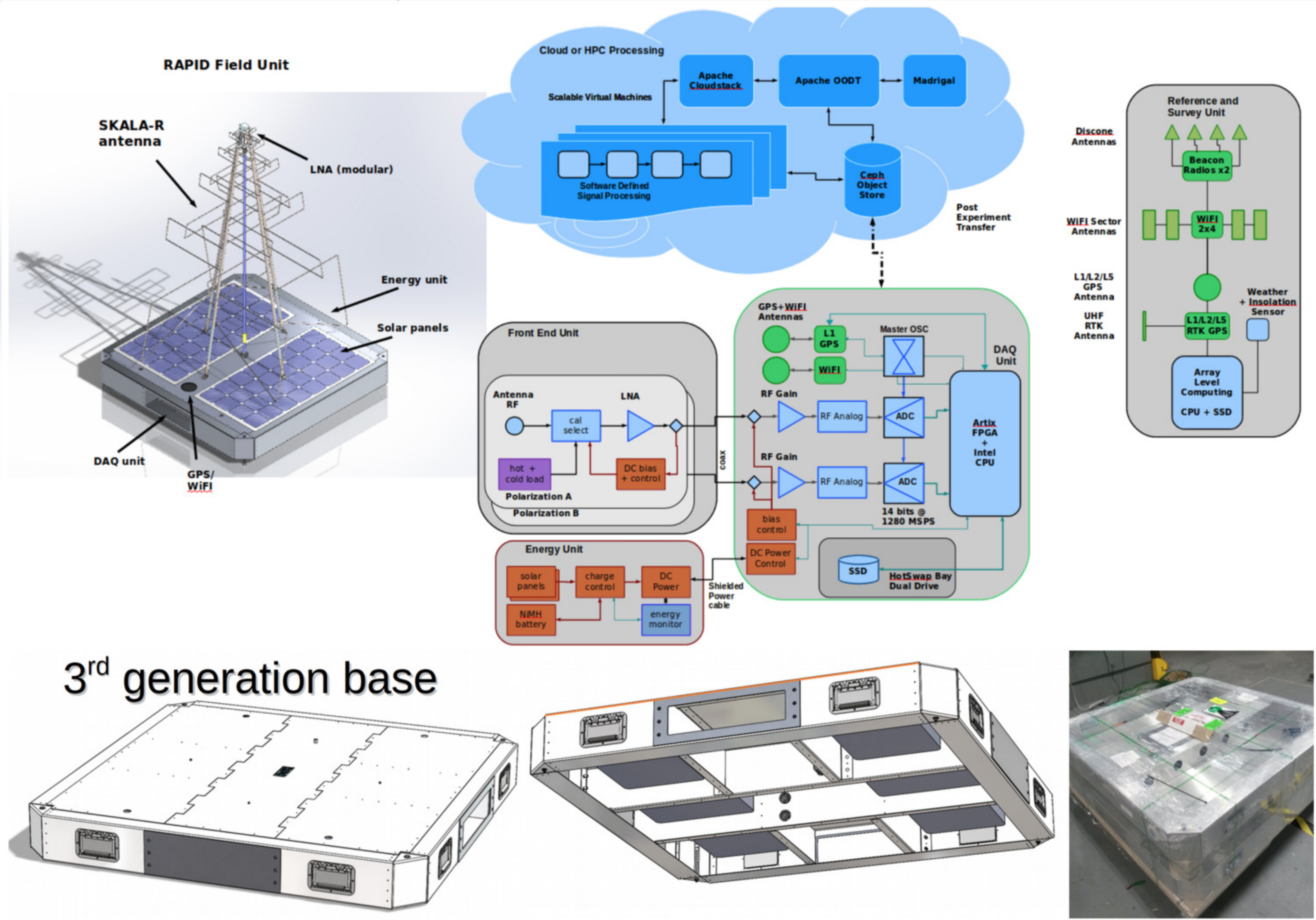}
\caption{Overview of the RAPID system design. At left is a complete, fully self-contained, portable and shippable RAPID element.  To the right is a high-level block diagram of the system, supporting direct RF sampling for the 50--600~MHz band, with recording to on-board solid state disks. Total power dissipation at full load is $\sim$30 Watts. Below left is the base unit, and below right are two RAPID units stacked and packaged on a pallet for shipping (to Peru in this instance).}
\label{fig3}
\end{figure}

RAPID uses a variant of the SKALA low frequency antenna developed at Cambridge, with a frequency range of $\sim$50--600~MHz, good sensitivity down to $\sim$30 degrees elevation, and a smooth power pattern that is not a strong function of frequency [de Lera Acedo et al., 2015]. A broader frequency range is accessible using different antennas, and the digital receiver system is capable of processing RF from 10~MHz to 2~GHz. A custom-built energy unit responds flexibly to changing battery and solar panel conditions, supports a variety of battery types, and is engineered for extremely low electro-magnetic interference (EMI). A highly integrated custom data acquisition (DAQ) unit supports wideband RF sampling, flexible data stream handling, and Gbyte/sec recording rates to hot-swappable SSD units.  A high quality crystal oscillator provides coherence between RAPID units, sufficient for phase-stable interferometry at all operating frequencies. Correlation is performed offline in a cloud computing environment. Operationally, the array will be deployed in campaign mode to locations and using physical configurations targeted at specific, narrowly focused scientific experiments. The initial scope of the RAPID system is envisaged as 50--70 elements. As of March 2017, the project was entering the production prototype phase for many key elements of the system. Figure~3 illustrates several aspects of the RAPID system, and a full description can be found in Lind et al. [2013; 2015].

\subsection{RAPID capabilities for solar studies}

The RAPID system is particularly well-suited to solar imaging studies because it has very wide instantaneous frequency coverage, and the array layout can be readily optimized for the scale sizes and field of view characteristics of solar emissions. For various types of solar burst emissions at the local plasma frequency and harmonics thereof, the wide frequency coverage corresponds to an ability to monitor a wide range of heights and electron densities in the corona simultaneously, and the customization of the array layout and extent offers the capability to make images and spot maps at resolutions limited only by the scattering properties of the coronal and heliospheric plasma, across the full frequency range. 

\begin{figure}[ht]
\centering
\includegraphics[width=0.6\textwidth]{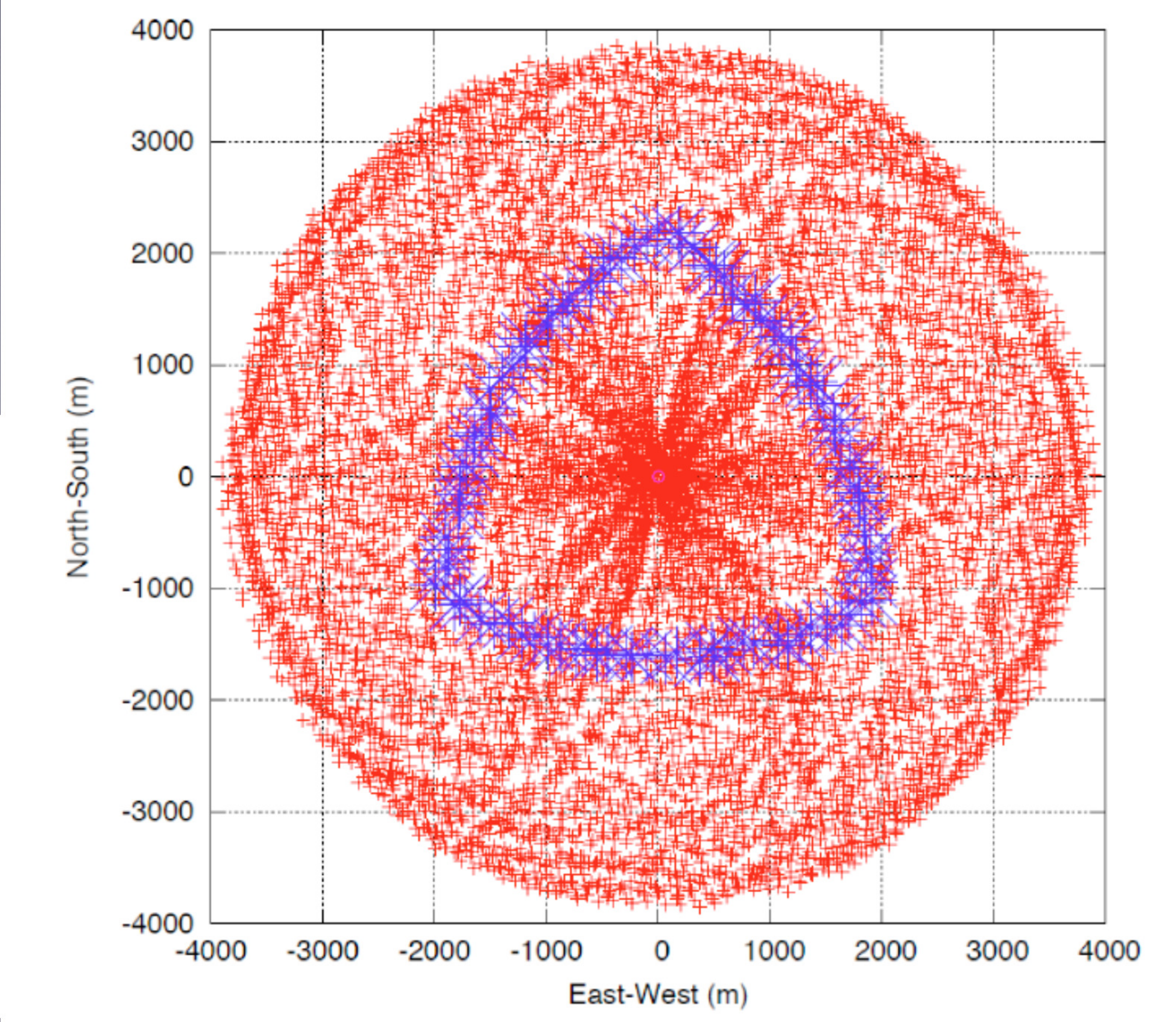}
\caption{Illustration of a possible RAPID array configuration, optimized for solar imaging applications.  In this example, 100 RAPID elements are deployed in a randomly dithered Reuleaux triangle of diameter ~4~km (blue symbols), yielding the instantaneous zenith $uv$ coverage shown by the red symbols. Such a configuration is capable of delivering excellent image fidelity, monochromatically and at high time resolution, across the full $\sim$1 degree angular extent of solar radio emissions in the 50--600~MHz frequency range.}
\label{fig4}
\end{figure}

Typically, fixed array layouts (like the MWA or LOFAR) must include a wide dynamic range of baseline lengths in a quest for adequate sampling of the $uv$ plane across many angular scale sizes  on the sky, thereby extending the scientific reach of the instrument.  Thus for any specific application (such as solar imaging), some potentially large fraction of the baselines are  either redundant (e.g. short baselines to which the target of interest appears as a point source) or insensitive (e.g. long baselines which resolve out all the structures of interest).  By contrast, the deployability and reconfigurability of RAPID allows a high degree of optimization of the layout to maximize the collection of scientifically useful information for a give number of antennas. For solar imaging, we are concerned  with structures no larger than $\sim$1 degree, so baselines shorter than a few tens of wavelengths are not required.  Similarly, and particularly at the higher frequencies of the 50--600~MHz RAPID range, quite compact structures can be anticipated, so an extended array configuration, with baselines of up to 10 or 20~km, is required to capture comprehensive information.

In Figure~4 we illustrate a notional RAPID layout, with 100 antennas, optimized for solar work, and spanning several km to support high resolution imaging. This layout is based on a randomly dithered Reuleaux triangle (blue) yielding extraordinarily uniform and dense sampling of the $uv$ plane (red). Using approaches such as this, the flexibility of RAPID can be exploited to provide much more efficient resource utilization for a wide range of scientific investigations. The nature of such a flexible and portable system means that it is best suited to applications involving relatively strong signals, because large collecting areas to pursue faint source science would be costly. For solar work, however, RAPID and instruments based on similar principles and technologies are poised to revolutionize the field. Solar imaging experiments with RAPID are planned as an early priority for the instrument commissioning phase.

\section{Conclusions}

Recent advances in low frequency array technology promise dramatic improvements in our ability to study the radio Sun in exquisite detail. By fully exploiting the high SNR that can be readily achieved in short integration times and at fine spectral resolution, generating detailed imagery for each instant and each frequency, low frequency radio studies can be expected to rise quickly in importance as a component of the multi-waveband toolbox for solar astronomy. The advent of capable low-frequency imaging arrays in diverse geographic locations will usher in an era of continuous monitoring of complex solar radio behaviors.

\textbf{Acknowledgements.} The authors gratefully acknowledge support from the US Air Force Office of Scientific Research (AFOSR) under Award No. FA9550-14-1-0192, and from the US National Science Foundation under award AGS-1229036. This work makes use of the Murchison Radio astronomy Observatory, operated by CSIRO. We acknowledge the Wajarri Yamatji people as the traditional owners of the Observatory site. Support for the operation of the MWA is provided by the Australian Government (NCRIS), under a contract to Curtin University administered by Astronomy Australia Limited. We acknowledge the Pawsey Super-computing Centre which is supported by the Western Australian and Australian Governments.

\section*{References}
\everypar={\hangindent=1truecm \hangafter=1}

Bastian, T.\,S., Synchrotron radio emission from a fast halo coronal mass ejection, \textsl{ApJ}, \textbf{665}, 805--812, 2007.

Benkevitch, L.\,V., A.\,E.\,E. Rogers, C.\,J. Lonsdale, R.\,J. Cappallo, D. Oberoi, P.\,J. Erickson, and K.\,A.\,V. Baker, Van Vleck correction generalization for complex correlators with multilevel quantization, arXiv:1608.04367, 2016.

Bowman, J.\,D., I.~Cairns, D.\,L.~Kaplan, T. Murphy, D. Oberoi, L. Staveley--Smith, W. Arcus, D.\,G. Barnes, G.~Bernardi, F.\,H. Briggs, S.~Brown, J.\,D.~Bunton, A.\,J.~Burgasser, R.\,J.~Cappallo, S.~Chatterjee, B.\,E.~Corey, A.~Coster, A. Deshpande, L.~deSouza, D. Emrich, P. Erickson, R.\,F.~Goeke, B.\,M.~Gaensler, L.\,J.~Greenhill, L. Harvey--Smith, B.\,J.~Hazelton, D.~Herne, J.\,N. Hewitt, M.~Johnston--Hollitt, J.\,C.~Kasper, B.\,B.~Kincaid, R.~Koenig, E. Kratzenberg, C.\,J.~Lonsdale, M.\,J.~Lynch, L.\,D.~Matthews, S.\,R.~McWhirter, D.\,A.~Mitchell, M.\,F.~Morales, E.\,H. Morgan, S.\,M.~Ord, J.~Pathikulangara, T.~Prabu, R.\,A. Remillard, T.~Robishaw, A.\,E.\,E. Rogers, A.\,A.~Roshi, J.\,E.~Salah, R.\,J.~Sault, N.\,U.~Shankar, K.\,S.~Srivani, J.\,B.~Stevens, R.~Subrahmanyan, S.\,J.~Tingay, R.\,B.~Wayth, M.~Waterson, R.\,L.~Webster, A.\,R.~Whitney, A.\,J.~Williams, C.\,L.~Williams, and J.\,S.\,B.~Wyithe, Science with the Murchison Widefield Array, \textsl{PASA}, \textbf{30}, 31--58, 2013.
    
de Lera Acedo, E., N. Razavi--Ghods, N. Troop, N. Drought, and A.\,J. Faulkner, SKALA, a log-periodic array antenna for the SKA-low instrument: design, simulations, tests and system considerations, \textsl{Experimental Astronomy}, \textbf{39}, 567--594, 2015.

Hurley--Walker, N., J.\,R. Callingham, P.\,J. Hancock, T.\,M.\,O. Franzen, L. Hindson, A.\,D. Kapi\'nska, J. Morgan, A.\,R. Offringa, R.\,B. Wayth, C. Wu, Q. Zheng, T. Murphy, M.\,E. Bell, K.\,S. Dwarakanath, B. For, B.\,M. Gaensler, M. Johnston--Hollitt, E. Lenc, P. Procopio, L. Staveley--Smith, R. Ekers, J.\,D. Bowman, F. Briggs, R.\,J. Cappallo, A.\,A. Deshpande, L. Greenhill, B.\,J. Hazelton, D.\,L. Kaplan, C.\,J. Lonsdale, S.\,R. McWhirter, D.\,A. Mitchell, M.\,F. Morales, E. Morgan, D. Oberoi, S.\,M. Ord, T. Prabu, N.\,U. Shankar, K.\,S. Srivani, R. Subrahmanyan, S.\,J. Tingay, R.\,L. Webster, A. Williams, and C.\,L. Williams,  GaLactic and extragalactic all-sky Murchison Widefield Array (GLEAM) survey - I. A low-frequency extragalactic catalogue, \textsl{MNRAS}, \textbf{464}, 1146--1167, 2017.

Lind, F.\,D., C.\,J. Lonsdale, A.\,J. Faulkner, P. Alexander, and C. Mattmann, Radio array of portable interferometric detectors (RAPID), \textsl{IEEE International Symposium on Phased Array Systems and Technology}, 676--683, 2013.

Lind, F.\,D., C.\,J. Lonsdale, A.\,J. Faulkner, C. Mattmann, N. Razavi--Ghods, E. de Lera Acedo, P. Alexander, J. Marchese, R. McWhirter, C. Eckert, J. Vierinen, R. Schaefer, W. Rideout, R. Cappallo, V. Pankratius, D. Oberoi, S. Khudikyan, M. Joyce, C. Goodale, M. Boustani, L. Cinquini, R. Verma, and M. Starch. Radio Array of Portable Interferometric Detectors (RAPID): Development of a deployable multiple application radio array, \textsl{2015 ICEAA}, 1337--1340, 2015.

Lonsdale, C.\,J., R.\,J. Cappallo, M.\,F. Morales, F.\,H. Briggs, L. Benkevitch, J.\,D. Bowman, J.\,D. Bunton, S. Burns, B.\,E.  Corey, L. Desouza, S.\,S. Doeleman, M. Derome, A. Deshpande, M.\,R. Gopala, L.\,J. Greenhill, D.\,E. Herne, J.\,N. Hewitt, P.\,A.  Kamini, J.\,C. Kasper, B.\,B. Kincaid, J. Kocz, E. Kowald, E. Kratzenberg, D. Kumar, M.\,J. Lynch, S. Madhavi, M. Matejek, D.\,A.  Mitchell, E. Morgan, D. Oberoi, S. Ord, J. Pathikulangara, T. Prabu, A. Rogers, A. Roshi, J.\,E. Salah, R.\,J. Sault, N.\,U. Shankar, K.\,S. Srivani, J. Stevens, S. Tingay, A. Vaccarella, M. Waterson, R.\,B. Wayth, R.\,L. Webster, A.\,R. Whitney, A.  Williams, and C. Williams, The Murchison Widefield Array: Design overview, \textsl{Proc. IEEE}, \textbf{97}, 1497--1506, 2009.

Smirnov, O., Revisiting the radio interferometer measurement equation. I. A full-sky Jones formalism, \textsl{A\&A}, \textbf{527}, 106--116, 2011.

Tingay, S.\,J., R. Goeke, J.\,D. Bowman, D. Emrich, S.\,M. Ord, D.\,A. Mitchell, M.\,F. Morales, T. Booler, B. Crosse, R.\,B. Wayth, C.\,J. Lonsdale, S. Tremblay, D. Pallot, T. Colegate, A. Wicenec, N. Kudryavtseva, W. Arcus, D. Barnes, G. Bernardi, F. Briggs, S. Burns, J.\,D. Bunton, R.\,J. Cappallo, B.\,E. Corey, A. Deshpande, L. Desouza, B.\,M. Gaensler, L.\,J. Greenhill, P.\,J. Hall, B.\,J. Hazelton, D. Herne, J.\,N. Hewitt, M. Johnston--Hollitt, D.\,L. Kaplan, J.\,C. Kasper, B.\,B. Kincaid, R. Koenig, E. Kratzenberg, M.\,J. Lynch, B. McKinley, S.\,R. McWhirter, E. Morgan, D. Oberoi, J. Pathikulangara, T. Prabu, R.\,A. Remillard, A.\,E.\,E. Rogers, A. Roshi, J.\,E. Salah, R.\,J. Sault, N. Udaya--Shankar, F. Schlagenhaufer, K.\,S. Srivani, J. Stevens, R. Subrahmanyan, M. Waterson, R.\,L. Webster, A.\,R. Whitney, A. Williams, C.\,L. Williams, and J.\,S.\,B. Wyithe, The Murchison Widefield Array: The Square Kilometre Array precursor at low radio frequencies, \textsl{PASA}, \textbf{30}, 7--27, 2013.

\end{document}